\documentclass{article}
\usepackage{spconf,amsmath,graphicx}
\usepackage{booktabs, multirow, hyperref, url}

\title{SponTTS: modeling and transferring spontaneous style for TTS}
\name{Hanzhao Li$^{1}$, Xinfa Zhu$^{1}$, Liumeng Xue$^{2}$, Yang Song, Yunlin Chen$^{3}$,  Lei Xie$^{1*}$\thanks{* Corresponding author.}}

\address{$^1$Audio, Speech and Language Processing Group (ASLP@NPU), School of Computer Science, \\ Northwestern Polytechnical University, Xi'an, China\\
$^2$ School of Data Science, The Chinese University of Hong Kong, Shenzhen (CUHK-Shenzhen), China\\
$^3$ Shanghai Mobvoi Information Technology Co., Ltd}
\begin{document}
\ninept
\maketitle
\begin{abstract}
Spontaneous speaking style exhibits notable differences from other speaking styles due to various spontaneous phenomena (e.g., filled pauses, prolongation) and substantial prosody variation (e.g., diverse pitch and duration variation, occasional non-verbal speech like a smile), posing challenges to modeling and prediction of spontaneous style. Moreover, the limitation of high-quality spontaneous data constrains spontaneous speech generation for speakers without spontaneous data. To address these problems, we propose \textit{SponTTS}, a two-stage approach based on neural bottleneck (BN) features to model and transfer spontaneous style for TTS. In the first stage, we adopt a Conditional Variational Autoencoder (CVAE) to capture spontaneous prosody from a BN feature and involve the spontaneous phenomena by the constraint of spontaneous phenomena embedding prediction loss. Besides, we introduce a flow-based predictor to predict a latent spontaneous style representation from the text, which enriches the prosody and context-specific spontaneous phenomena during inference. In the second stage, we adopt a VITS-like module to transfer the spontaneous style learned in the first stage to the target speakers. Experiments demonstrate that SponTTS is effective in modeling spontaneous style and transferring the style to the target speakers, generating spontaneous speech with high naturalness, expressiveness, and speaker similarity. The zero-shot spontaneous style TTS test further verifies the generalization and robustness of SponTTS in generating spontaneous speech for unseen speakers.

\end{abstract}
\section{Introduction}
\label{sec:intro}

One of the primary goals of speech synthesis or text-to-speech (TTS)  is to produce speech that sounds as natural and human-like as possible. Spontaneous-style speech synthesis aims to replicate how humans naturally speak, including tone, pace, pitch variations, and various spontaneous events. With the rapid advancements in deep learning, speech synthesis has achieved remarkable progress in synthesizing highly natural speech, approaching the quality that is even nearly indistinguishable from human speech~\cite{DBLP:journals/corr/abs-2106-15561, DBLP:conf/iclr/0006H0QZZL21, DBLP:conf/nips/KongKB20, DBLP:conf/icml/KimKS21}. However, existing TTS systems cannot perform sufficiently in generating spontaneous-style speech.

To deliver stylistic speech, a typical speech synthesis system usually adopts a reference encoder to capture prosodic variation in a reference audio~\cite{Wang2018StyleTU}, where the reference audio is always required to capture the desired speaking style. However, it is impractical to find a suitable audio reference for spontaneous speech synthesis because spontaneous speech is often characterized by context-specific \textit{spontaneous phenomena}~\cite{DBLP:conf/slsp/QaderLLS18,DBLP:conf/interspeech/SzekelyHBG19a, DBLP:conf/slt/GuoZSHX21}, such as filled pauses, changes in speaking rate, and stress, etc.



Using explicit phenomena labels to model spontaneous style is a typical solution~\cite{DBLP:conf/interspeech/CongYHLX021, DBLP:journals/corr/abs-2107-02530}. Although those methods demonstrate the feasibility and controllability of spontaneous phenomena for spontaneous speech generation, they neglect the intrinsic prosody rendering in spontaneous speech, which can not be labeled accurately, such as diverse pitch and duration variation and occasional non-verbal speech, resulting in limited expressivity in the synthetic speech. A few studies~\cite{Shechtman2021AcquiringCS} manually compute prosody statistics to model spontaneous prosody, but spontaneous phenomena are not considered, causing the lack of natural spontaneous phenomena expression in the synthesized speech. 


Moreover, previous studies have reported that spontaneous speech exhibits notable differences compared to reading-style speech due to disfluency and substantial prosody variation~\cite{DBLP:conf/icassp/SzekelyHBG20, Winkworth1994VariabilityAC, DBLP:conf/interspeech/SzekelyMG17}, which pose a significant challenge when predicting the spontaneous style representation from the text during inference. Some works~\cite{DBLP:conf/interspeech/CongYHLX021,DBLP:journals/corr/abs-2107-02530} adopt an additional text-based predictor for spontaneous phenomena label prediction, which may suffer from unnatural generated speech when phenomena labels are present unreasonably. Furthermore, although it is possible to synthesize spontaneous speech with diverse spontaneous phenomena by manually specifying spontaneous phenomenon labels during inference, it is not practical in real applications. A recent work~\cite{Li2023TowardsSS} predicts \textit{soft} phenomenon label embedding, jointly trained with the TTS model, to reduce the cumulation of prediction errors in the generated speech. Unlike these works, in this paper, we predict a latent spontaneous style representation that involves richer spontaneous prosody and more types of spontaneous phenomena.



Additionally, spontaneous style transfer, aiming to synthesize spontaneous-style speech for speakers who do not have spontaneous style data, is a practical way to alleviate data limitation and enrich speech expressivity. However, it is challenging due to specific textual patterns in spontaneous speech and the entangled correlation between spontaneous style and speaker identity. Some previous work~\cite{Fernandez2022TransplantationOC} explores the leverage of Voice-Conversion-based data augmentation in spontaneous style transfer but it is constrained by the quality of the voice conversion system. Bottleneck (BN) features extracted from a well-trained automatic speech recognition (ASR) model are considered to be linguistic- and style-rich but speaker-independent and widely used for emotion and style transfer in TTS~\cite{DBLP:journals/corr/abs-2211-10568, Liu2021RefereeTR, DBLP:conf/icassp/HuangXYMQ21}, which inspires us to use the BN features for spontaneous style modeling and transferring.

This paper proposes a novel TTS approach to model and transfer spontaneous style for TTS, named \textit{SponTTS}, a two-stage framework based on BN features. In the first stage, we adopt a Conditional Variational Autoencoder (CVAE) to learn a latent representation of spontaneous style, which captures spontaneous prosody from the BN feature and delivers spontaneous phenomena by the constraint of spontaneous phenomena embedding prediction loss. Moreover, we utilize a flow-based predictor to predict the latent representation from the text, which enriches the prosody and context-specific phenomena during inference. In the second stage, we adopt a VITS-like module to transfer the spontaneous style learned in the first stage to a target speaker, either seen or unseen in the training dataset with only reading-style data. Objective and subjective experiments demonstrate that SponTTS effectively models the spontaneous style of speech and successfully transfers it to a target speaker, generating spontaneous speech with high naturalness, expressiveness, and speaker similarity. The zero-shot spontaneous style TTS test further verifies the generalization and robustness of SponTTS in generating spontaneous-style speech for unseen speakers. We suggest the readers listen to our online demos\footnote{Demo:\href{https://kkksuper.github.io/SponTTS/}{https://kkksuper.github.io/SponTTS/}}.

\section{SponTTS}
\label{sec:format}

\begin{figure}[htbp]
    \centering
    \vspace{-10pt}
    \centerline{\includegraphics[width=8.0cm]{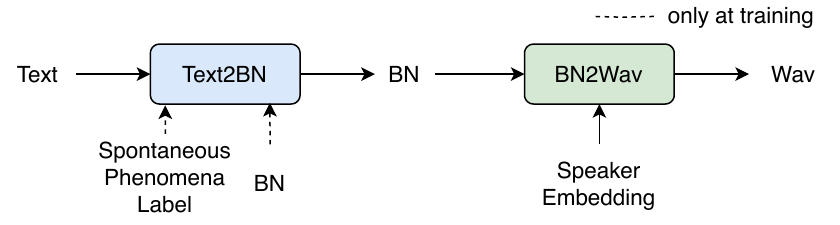}}
  
    \caption{The overview of SponTTS}
    \label{fig:overall}
\end{figure}

As shown in Figure~\ref{fig:overall}, based on the BN features, we train SponTTS in two stages. The first stage is Text-to-BN (Text2BN) which is designed to map text to speaker-independent BN features under the condition of spontaneous style representation $z$ learned from both BN and spontaneous phenomena label, rendering spontaneous prosody and spontaneous phenomena simultaneously. The second stage is BN-to-Wave (BN2Wave) which aims to directly generate the waveform from BN conditioned on speaker embedding, achieving spontaneous-style speech synthesis for anyone unseen in training.



\vspace{-20pt}

\subsection{Text-to-BN module}
\label{ssec:subhead}
\begin{figure}[htbp]
    \centering
    \centerline{\includegraphics[width=7.5cm]{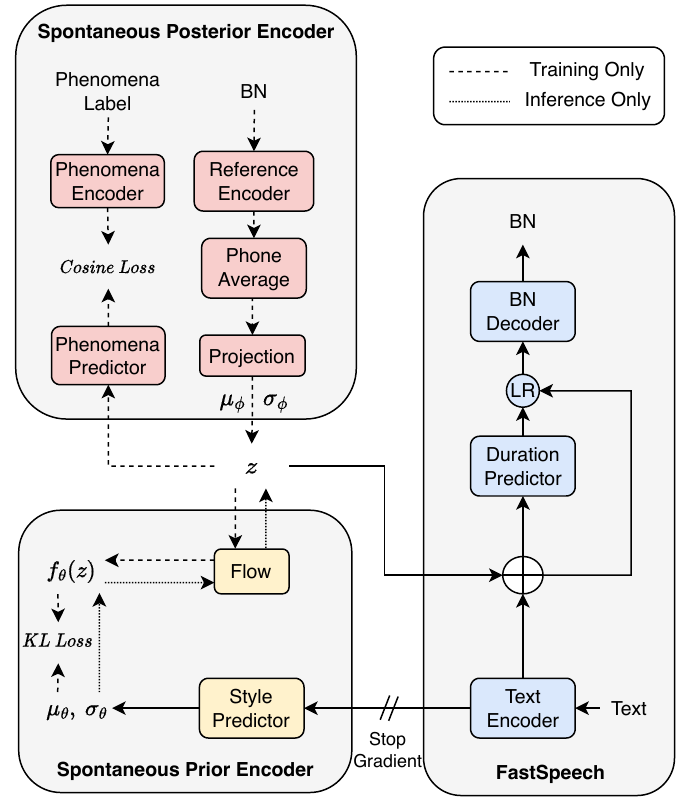}}
    \vspace{-3pt}
    \caption{The structure of Text2BN}  \vspace{-8pt}
    \label{fig:text2bn}
\end{figure}

The structure of Text2BN is presented in Figure~\ref{fig:text2bn}, which follows  FastSpeech~\cite{DBLP:conf/nips/RenRTQZZL19} as the backbone to map text to BN. It mainly comprises three key components: a text encoder, a length regulator, and a BN decoder. The text encoder takes the text as input and produces phoneme-level encoded output, which is then expanded to the frame level by the length regulator according to the duration predicted from the duration predictor and consumed by the BN decoder to generate the BN features. To model spontaneous style, we propose a CVAE structure to obtain the posterior distribution $q_{\phi}(z|x_{bn})$ from BN features $x_{bn}$ in a spontaneous posterior encoder. We combine $z$ with the text encoder output via an addition operation for BN prediction. Furthermore, a flow-based spontaneous prior encoder jointly optimized with the backbone and spontaneous posterior encoder is proposed to acquire the utterance-specific prior distribution and then sample spontaneous style representation $z$ from the prior distribution during inference.

\textbf{Spontaneous Posterior Encoder} We design a spontaneous posterior encoder to capture the various spontaneous phenomena and diverse spontaneous prosody. We first adopt a reference encoder to learn a frame-level representation from the BN features extracted from target speech $x_{bn}$ and perform phoneme-level average pooling according to the durations obtained through an HMM-based force alignment model to downsample the representation to the phoneme level. Subsequently, we employ the reparametrization trick~\cite{DBLP:journals/corr/KingmaW13} to get a spontaneous style latent representation $z$ from the mean $\mu_{\phi}$ and variance $\sigma_{\phi}$ leaned from a projection layer. To encourage the posterior distribution $q_\phi(z|x_{bn})$ associated with the knowledge of spontaneous phenomenons, we introduce a phenomena predictor as a constraint on $z$ to minimize the distance between the predicted spontaneous phenomena embedding from $z$ and ground-truth spontaneous phenomena embedding obtained from spontaneous phenomena labels. This constraint method can avoid the instability problem when modeling phenomena using separate classifiers for different categories of spontaneous phenomena. The posterior distribution can be formulated as:
\vspace{-3pt}
\begin{equation}
    \begin{aligned}
        z \sim q_{\phi}(z|x_{bn})=N(z;\mu_{\phi}(x_{bn}),\sigma_{\phi}(x_{bn})).
    \end{aligned}
\end{equation}
\vspace{-13pt}

\textbf{Spontaneous Prior Encoder} We introduce a spontaneous prior encoder to obtain a prior distribution $N(\mu_{\theta}(c), \sigma_{\theta}(c))$ from the text encoder output $c$, as shown in the left bottom part of Figure~\ref{fig:text2bn}. Inspired by VITS~\cite{DBLP:conf/icml/KimKS21}, we use a normalized flow $f_{\theta}$ to map the prior distribution $p(f_\theta(z)|c)$ to a more complex distribution, enhancing prosody variance predicted from the text during inference:
\vspace{-3pt}
\begin{equation}
    \begin{aligned}
        p_{\theta}(z|c) = N(f_\theta(z);\mu_{\theta}(c),\sigma_{\theta}(c))\left | \det \frac{\partial f_\theta(z)}{\partial z}\right |
    \end{aligned}
\end{equation}
\vspace{-10pt}



The training objective of the Text2BN module $L_{t2bn}$ is:
\vspace{-3pt}
\begin{equation}
    L_{t2bn} = L_{BN} + L_{dur} + L_{phen} + \alpha \cdot L_{style}
\end{equation}
\vspace{-10pt}
\begin{equation}
    \begin{aligned}
        L_{style}=\log q_{\phi}(z|x_{bn})-\log p_{\theta}(z|c)
    \end{aligned}
\end{equation}
where $L_{BN}$ is the L2 reconstruction loss between predicted BN and ground-truth BN; $L_{dur}$ is the L2 duration prediction loss. $L_{phen}$ is the cosine distance loss for phenomena prediction; $L_{style}$ is the Kullback–Leibler divergence (KLD) used for spontaneous prosody prediction, where we use $\alpha$ to anneal the KLD loss to resolve KL collapse problem.


\subsection{BN-to-Wave module}
\label{ssec:subhead}
\begin{figure}[htbp]
    \centering
    \centerline{\includegraphics[width=7.0cm]{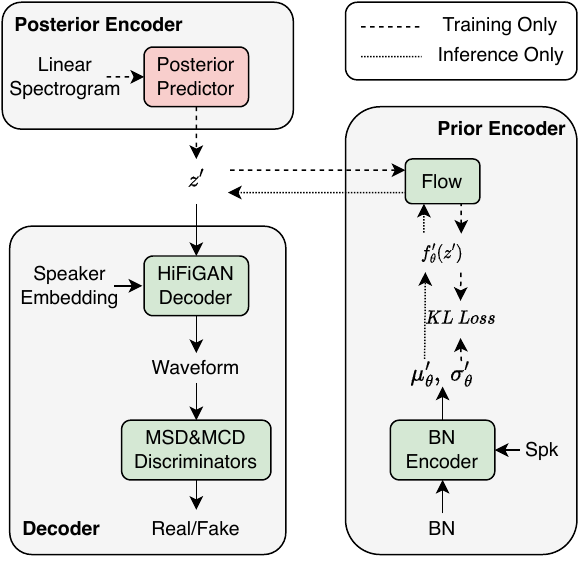}}
    \vspace{-3pt}
    \caption{The structure of BN2Wave}
    \label{fig:bn2wav}
\end{figure}

We build the BN2Wave following VITS~\cite{DBLP:conf/icml/KimKS21}, which is composed of a prior encoder, posterior encoder and decoder, as presented in Figure~\ref{fig:bn2wav}.
\vspace{-3pt}
\vspace{-3pt}
In the prior encoder, the BN encoder and flow generate the prior distribution $p_{\theta}(z^ {\prime}|c^ {\prime})$ of the CVAE, where $c^ {\prime}$ is the condition information, including the BN features $x_{bn}$ and speaker embedding $spk$. The posterior encoder mainly composes a posterior predictor, which takes a linear spectrum $x_{lin}$ as input and produces the posterior distribution $q(z^ {\prime}|x_{lin})$. The posterior predictor comprises a stack of WaveNet residual blocks~\cite{DBLP:conf/ssw/OordDZSVGKSK16}. Finally, the HIFI-GAN~\cite{DBLP:conf/nips/KongKB20} decoder generates the audible waveform from $z^\prime$ conditioned on the speaker embedding. We leverage a pre-trained speaker verification model rather than a look-up table to obtain more stable and robust speaker embeddings, expecting to achieve speaker-adaptive spontaneous speech synthesis. 
\vspace{-3pt}
\begin{equation}
    \begin{aligned}
        z^ {\prime} \sim q_{\phi}(z^ {\prime}|x_{lin})=N(z^ {\prime};\mu^ {\prime}_{\phi}(x_{lin}),\sigma^ {\prime}_{\phi}(x_{lin})).
    \end{aligned}
\end{equation}
\vspace{-10pt}
\begin{equation}
    \begin{aligned}
        p_{\theta}(z^ {\prime}|c^ {\prime}) = N(f^ {\prime}_\theta(z^ {\prime});\mu^ {\prime}_{\theta}(c^ {\prime}),\sigma^ {\prime}_{\theta}(c^ {\prime}))\left | \det \frac{\partial f^ {\prime}_\theta(z^ {\prime})}{\partial z^ {\prime}}\right |
    \end{aligned}
\end{equation}
\begin{equation}
    \begin{aligned}
        c^{\prime} = [x_{bn}, spk]
    \end{aligned}
\end{equation}
\vspace{-10pt}

We utilize the L1 loss as the reconstruction loss $L_{rec}$ to minimize the distance between the mel-spectrograms of the ground-truth waveform and the predicted waveform. Furthermore, similar to VITS~\cite{DBLP:conf/icml/KimKS21}, the decoder integrates the adversarial training loss $L_{adv}$ and feature map loss $L_{fm}$ to improve the performance of waveform generation further. The training objectives of the BN2Wave module $L_{bn2w}$ is
\vspace{-3pt}
\begin{equation}
    L_{bn2w}^{G} = L_{kl} + 45 \cdot L_{rec} + L_{adv}(G) + L_{fm}(G)
\end{equation}
\vspace{-10pt}
\begin{equation}
    L_{bn2w}^{D} = L_{adv}(D)
\end{equation}
\vspace{-10pt}
\begin{equation}
    L_{kl}=\log q_{\phi}(z^{\prime}|x_{lin})-\log p_{\theta}(z^{\prime}|c^{\prime})
\end{equation}
where $L_{bn2w}^{G}$, $L_{bn2w}^{D}$, and $L_{kl}$ are the generative loss, discriminator loss, and KL divergence of the hidden distribution. Same as HiFi-GAN~\cite{DBLP:conf/nips/KongKB20}, we set the loss rate of mel-spectrum to 45.

\section{Experiments and Results}
\label{sec:pagestyle}

\subsection{Dataset}
\label{ssec:subhead}

We train the Text2BN module using an internal Mandarin spontaneous speech dataset. This dataset comprises 16.7 hours of spontaneous speech data from a female speaker engaged in casual conversations with other speakers. It contains five types of spontaneous phenomena annotations, including silent pauses, prolongation, fast speaking rate, liaison, and stress, where silent pauses, caused by thinking or hesitation, typically occur after any phoneme in an utterance, and stress is further divided into primary stress, secondary stress, and no stress. 


For BN2Wave module training, we also utilize a multi-speaker reading-style internal Mandarin dataset in addition to the dataset used in the Text2BN module. This dataset includes 340 speakers, each contributing between 200 and 1000 utterances, resulting in a total duration of 294.1 hours. 

In the experiments, we evaluate the performance of the proposed model in spontaneous style 
speech synthesis for both seen speakers and unseen speakers who only have reading-style speech data. We randomly select four speakers from the dataset used in BN2Wave module training for seen speakers. For unseen speakers, we also select four speakers from another reading-style dataset. Both the seen and unseen speakers include two males and two females.


\subsection{Model parameters and training details}
\label{ssec:subhead}

The text encoder and a BN decoder in the Text2BN module both consist of a 6-layer FFT block with a hidden dimension of 256. The reference encoder utilizes a WaveNet~\cite{DBLP:conf/ssw/OordDZSVGKSK16} structure with five layers and a hidden dimension of 192. The prosody representation $z$ has a dimension of 16. The phenomena encoder and phenomena predictor employs five linear layers, while the style predictor utilizes five convolutional layers with a kernel size of 5. The BN2Wave module follows the settings of VITS, and the BN encoder consists of 6 FFT blocks with a 192-dimensional hidden size. 

All the audio recordings are downsampled to 16kHz. We adopt ECAPA-TDNN~\cite{DBLP:conf/interspeech/DesplanquesTD20} to extract speaker embedding, which is trained on voxceleb with 351 hours and 1251 speakers~\cite{Nagrani2017VoxCelebAL}. Additionally, we adopt the WeNet U2++ model~\cite{DBLP:conf/interspeech/YaoWWZYYPCXL21} trained on WenetSpeech including 10,000 hours~\cite{DBLP:conf/icassp/ZhangLGSYXXBCZW22} as the ASR model to extract the BN features with the dimension of 512. The BN features are further interpolated to 50ms frame length and 12.5 ms frameshift for better TTS performance.

\subsection{Comparison Models}
\label{ssec:subhead}

We compare our proposed model \textbf{SponTTS} with the following three models which are built based on SponTTS for fair comparison:

\begin{itemize}
    \item [$\bullet$] \textbf{Baseline}: We remove the spontaneous posterior encoder and prior encoder from SponTTS, a vanilla two-stage TTS framework without spontaneous style modeling.
    \item [$\bullet$] \textbf{TP}: We remove the reference encoder and variational processes, and the style predictor directly predicts phenomena embedding corresponding to the output of the phenomena encoder, which means only spontaneous phenomena modeling is considered.

    \item [$\bullet$] \textbf{TPVAE}: We remove the phenomena predictor and phenomena encoder from SponTTS, which means we only model spontaneous prosody. 
\end{itemize}

\subsection{Objective Evaluation}  \vspace{-2pt}
\label{ssec:subhead}

To investigate the difference in prosodic variability between the reading-style and spontaneous-style speech, we first calculate the standard deviations of the F0 (F0 std) and duration (Dur. std) on all speech data of a female speaker in the reading-style corpus and spontaneous-style corpus used in BN2Wave module training, respectively. We also compute F0 std and Dur. std on the 100 synthesized speech utterances with the speaker identity of the female from all four models.
We use the harvest~\cite{DBLP:conf/interspeech/Morise17} to extract the F0 and an HMM-based force alignment model to obtain the duration.
Furthermore, we calculate the character error rate (CER) on the synthesized speech by using a pre-trained WeNet~\cite{Yao2021WeNetPO} to measure the robustness of the models. 

The results listed in Table \ref{variance} show that spontaneous speech exhibits a higher standard deviation of F0 and duration than reading-style speech, indicating spontaneous speech delivers more varied prosody than reading-style speech. When comparing the results from different models, SponTTS obtains the highest standard deviation for F0 and duration and the lowest in CER, indicating that synthetic speech from SponTTS has richer rhythmic variability and more stable naturalness. Furthermore, TP outperforms Baseline and TPVAE in the standard deviation of F0, duration, and CER, suggesting that spontaneous phenomena improve the prosody variation of spontaneous style and model stability. In addition, TPVAE gets lower variance in F0 and duration but higher CER, suggesting the importance of incorporating spontaneous phenomena modeling in spontaneous style TTS.

We also calculate the cosine similarity (S-Cosine) to verify the speaker similarity further. To this end, we extract a 256-dimensional speaker embedding via the Resemblyzer tool \cite{DBLP:conf/nips/JiaZWWSRCNPLW18, DBLP:conf/icassp/WanWPL18} from the recordings and synthesized speech. The results are listed in the last column of Table \ref{seen}. As we can see, all four models achieve comparable results, yielding above 0.8 cosine similarity. Specifically, SponTTS achieves slightly higher cosine similarity than other models. Objective results show that the proposed SponTTS captures various prosody while maintaining good speaker similarity.


\vspace{-17pt}

\begin{table}[h]
    \centering
    \caption{Results of prosody analysis.}
    \label{variance}
    \vspace{3pt}
\begin{tabular}{cccc}
\toprule
\multicolumn{1}{c}{\textbf{Method}} & \textbf{F0 std $\uparrow$}   & \textbf{Dur. std $\uparrow$} & \textbf{CER(\%) $\downarrow$}\\  \midrule
Reading Corpus       &    85.46      &     3.93    &    1.2     \\ 
Spontaneous Corpus   &    93.96      &     5.76    &    1.8      \\ \midrule
Baseline             &    91.07      &     4.81    &    6.0     \\ 
TP                   &    91.35      &     4.88    &    3.6     \\
TPVAE                &    88.89      &     4.31    &    4.1     \\
SponTTS              &   \textbf{92.18}  &  \textbf{4.98}  &    \textbf{2.6}     \\
\bottomrule
\end{tabular}
\end{table}

\vspace{-20pt}


\subsection{Subjective Evaluation}  \vspace{-2pt}
\label{ssec:subhead}
We conduct Mean Opinion Score (MOS) tests to evaluate speech naturalness (N-MOS) and speaker similarity (S-MOS) in generated spontaneous-style speech. We also perform comparison MOS (CMOS) and preference tests regarding spontaneous style, where participants are asked to select which speech is more spontaneous. The subjective testing set contains 25 utterances, and 20 native speakers participate in each test.

Table \ref{seen} shows the N-MOS and S-MOS results. SponTTS outperforms the other models in both speech naturalness and speaker similarity. Compared to the Baseline, the experimental results of TP demonstrate that spontaneous phenomena modeling improves naturalness, which is consistent with the objective result. Moreover, with the help of the posterior encoder that captures prosodic changes in spontaneous speech, TPVAE achieves perceptual improvements in naturalness compared to Baseline and TP. SponTTS achieves the highest naturalness and speaker similarity, showing that modeling spontaneous phenomena and prosody simultaneously in TTS effectively generates spontaneous-style speech with higher naturalness and speaker similarity.



\vspace{-15pt}

\begin{table}[h]
    \centering
    \caption{Results of subjective evaluation with 95\% confidence interval and speaker cosine similarity in seen speakers.}
    \label{seen}
    \vspace{3pt}
\begin{tabular}{cccc}
\toprule
\multicolumn{1}{c}{\textbf{Model}} & \textbf{N-MOS $\uparrow$}   & \textbf{S-MOS $\uparrow$} & \textbf{S-Cosine $\uparrow$} \\  \midrule
Baseline     &    3.61 $\pm$ 0.09   &    3.97 $\pm$ 0.05    &  0.818  \\
TP           &    3.70 $\pm$ 0.08   &    3.91 $\pm$ 0.07    &  0.835  \\
TPVAE        &    3.93 $\pm$ 0.09   &    3.99 $\pm$ 0.07    &  0.824  \\
SponTTS      &    \textbf{4.02 $\pm$ 0.08}   &    \textbf{4.06 $\pm$ 0.06}    &  \textbf{0.843}  \\
\bottomrule
\end{tabular}
\end{table}

\vspace{-5pt}

The CMOS and preference test results presented in Table \ref{cmos} show that SponTTS is preferable in spontaneous style (p-value $<$ 0.05). Compared to SponTTS, TP has a smaller gap than the other models, which shows the effectiveness of explicitly modeling spontaneous phenomena. 
Meanwhile, SponTTS outperforms TP, indicating that incorporating spontaneous prosody can further enhance the expressiveness in synthesized speech. In addition, the CMOS between TPVAE and SponTTS shows that the constraint of spontaneous phenomena on spontaneous representation helps to capture spontaneous style-related features, producing more expressive speech.


\vspace{-17pt}

\begin{table}[h]
\setlength\tabcolsep{3pt}
    \centering
    \caption{Results of spontaneous style preference test.}
    \label{cmos}
    \vspace{3pt}
\begin{tabular}{ccccc}
\toprule
           & \multirow{2}{*}{\textbf{CMOS} $\uparrow$} & \multicolumn{3}{c}{\textbf{Preference(\%)}} \\ \cmidrule{3-5} 
                     &                       & \textbf{Left}     & \textbf{Neutral}     & \textbf{Right}     \\ \midrule
Baseline vs. SponTTS &    0.56    &    22.2    &     22.2    &    55.6    \\
TP vs. SponTTS       &    0.25    &    36.1    &     13.7    &    50.2    \\
TPVAE vs. SponTTS    &    0.66    &    25.2    &     12.5    &    62.3    \\ \bottomrule
\end{tabular}
\end{table}

\vspace{-15pt}

\subsection{Zero-shot Spontaneous Style TTS} \vspace{-2pt}
We further evaluate our proposed SponTTS on transferring spontaneous style to unseen speakers to achieve zero-shot spontaneous style TTS. The N-MOS, S-MOS, and speaker cosine similarity are listed in Table~\ref{zeroshot}. As we can see, SponTTS performs exceptionally well for unseen speakers, with no significant performance degradation compared to the seen speakers. It indicates that our proposed method is robust to transfer spontaneous style to anyone who only has reading-style data and is unseen during training.

\vspace{-16pt}
\begin{table}[h]
    \centering
    \caption{Results of subjective evaluation with 95\% confidence interval and speaker cosine similarity in unseen speakers.}
    \label{zeroshot}
    \vspace{3pt}
\begin{tabular}{cccc}
\toprule
\multicolumn{1}{c}{\textbf{Model}} & \textbf{N-MOS $\uparrow$}   & \textbf{S-MOS $\uparrow$} & \textbf{S-Cosine $\uparrow$} \\  \midrule
Baseline     &    3.74 $\pm$ 0.09   &    3.72 $\pm$ 0.06    &  0.791  \\
TP           &    3.86 $\pm$ 0.09   &    3.68 $\pm$ 0.08    &  0.800  \\
TPVAE        &    3.83 $\pm$ 0.10   &    3.71 $\pm$ 0.09    &  0.802  \\
SponTTS      &    \textbf{3.94 $\pm$ 0.08}   &    \textbf{3.77 $\pm$  0.08}    &  \textbf{0.808}  \\
\bottomrule
\end{tabular}
\end{table}
\vspace{-16pt}

\section{Conclusion} \vspace{-5pt}
\label{sec:pagestyle}

This paper proposes SponTTS to model and transfer spontaneous style in TTS. Specifically, we adopt a CVAE to capture diverse spontaneous prosody and learn spontaneous phenomena simultaneously. We further utilize a flow-based predictor to enrich the prosody and context-specific phenomena predicted from the text. Objective and subjective experiments demonstrate that SponTTS effectively models spontaneous style and successfully transfers the spontaneous speaking style to seen and unseen speakers, generating natural and expressive speech with high speaker similarity.


\clearpage
\newpage

\bibliographystyle{IEEEbib}
\bibliography{refs}

\end{document}